\documentclass[sigconf]{acmart}
\usepackage[a-1b]{pdfx}

\copyrightyear{2023}
\acmYear{2023}
\setcopyright{rightsretained}
\acmConference[SIGIR '23]{Proceedings of the 46th International ACM SIGIR Conference on Research and Development in Information Retrieval}{July 23--27, 2023}{Taipei, Taiwan}
\acmBooktitle{Proceedings of the 46th International ACM SIGIR Conference on Research and Development in Information Retrieval (SIGIR '23), July 23--27, 2023, Taipei, Taiwan}\acmDOI{10.1145/3539618.3591740}
\acmISBN{978-1-4503-9408-6/23/07}

\usepackage{etoolbox}
\makeatletter
\patchcmd{\maketitle}{\@copyrightpermission}{
   \begin{minipage}{0.3\columnwidth}
     \href{http://creativecommons.org/licenses/by/4.0/}{\includegraphics[width=0.90\textwidth]{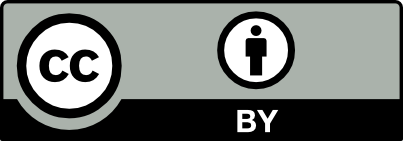}}
   \end{minipage}\hfill
   \begin{minipage}{0.7\columnwidth}
     \href{http://creativecommons.org/licenses/by/4.0/}{This work is licensed under a Creative Commons Attribution International 4.0 License.}
   \end{minipage}

   \vspace{5pt}
}{}{}

\usepackage{balance}

\makeatother

\usepackage{enumitem}
\newlist{inlinelist}{enumerate*}{1}
\setlist*[inlinelist,1]{%
  label=(\roman*),
}
\usepackage{subfigure}

\usepackage{xspace}
\usepackage{mathtools}
\usepackage{amsbsy}
\usepackage{multirow}
\usepackage{bbm}
\usepackage{siunitx}  

\DeclarePairedDelimiterX{\infdivx}[2]{(}{)}{%
  #1\;\delimsize\|\;#2%
}
\newcommand{\infdiv}{\text{KLD}_k\infdivx}
\newcommand{\tr}{\text{tr}}

\newcommand{\framework}{MRL\xspace}

\title{Multivariate Representation Learning for Information Retrieval}

\author{Hamed Zamani}
\affiliation{\institution{University of Massachusetts Amherst}
\country{United States}}
\email{zamani@cs.umass.edu}

\author{Michael Bendersky}
\affiliation{\institution{Google Research}
\country{United States}}
\email{bemike@google.com}

\begin{document}


\begin{abstract}
Dense retrieval models use bi-encoder network architectures for learning query and document representations. These representations are often in the form of a vector representation and their similarities are often computed using the dot product function. In this paper, we propose a new representation learning framework for dense retrieval. Instead of learning a vector for each query and document, our framework learns a multivariate distribution and uses negative multivariate KL divergence to compute the similarity between distributions. For simplicity and efficiency reasons, we assume that the distributions are multivariate normals and then train large language models to produce mean and variance vectors for these distributions. We provide a theoretical foundation for the proposed framework and show that it can be seamlessly integrated into the existing approximate nearest neighbor algorithms to perform retrieval efficiently. We conduct an extensive suite of experiments on a wide range of datasets, and demonstrate significant improvements compared to competitive dense retrieval models.
\end{abstract}



\maketitle

\section{Introduction}
\label{sec:intro}
Utilizing implicit or explicit relevance labels to learn retrieval models, also called learning-to-rank models, is at the core of information retrieval research. Due to efficiency and even sometimes effectiveness reservations, learning-to-rank models have been mostly used for \emph{reranking} documents retrieved by an efficient retrieval model, such as BM25 \cite{Robertson1995OkapiBM25}. Therefore, the performance of learning-to-rank models was bounded by the quality of candidate documents selected for reranking. In 2018, the SNRM model \cite{SNRM} has revolutionized the way we look at learning-to-rank models by arguing that bi-encoder neural networks can be used for representing queries and documents, and document representations can be then indexed for efficient retrieval at query time. The model applied learned latent sparse representations for queries and documents, and indexed the document representations using an inverted index. In 2020, the DPR model \cite{DPR} demonstrated that even bi-encoder networks with dense representations can be used for efficient retrieval. They took advantage of approximate nearest neighbor algorithms for indexing dense document representations. This category of models, often called \emph{dense retrieval} models, has attracted much attention and led to state-of-the-art performance on a wide range of retrieval tasks \cite{Khattab2020ColBERTEA,Xiong2021ApproximateNN,Zeng:2022:CLDRD,Qu2021RocketQAAO,Hofsttter2021EfficientlyTA}.

Existing sparse and dense representation learning models can be seen as instantiations of Salton et al.'s vector space models \cite{VSM}, i.e., queries and documents are represented using vectors and relevance is defined using vector similarity functions, such as inner product or cosine similarity. Such approaches suffer from a major shortcoming: they do not represent the model's \emph{confidence} on the learned representations. Inspired by prior work on modeling uncertainty in information retrieval (e.g., \cite{Cohen+al:2021, collins+callan:2007,Wang+Zhu:2009}), this paper builds upon the following hypothesis:
\begin{quote}
\emph{Neural retrieval models would benefit from modeling uncertainty (or confidence) in the learned query and document representations.}
\end{quote}

\begin{figure*}[t]
    \centering
    \subfigure[Representation learning in existing single-vector dense retrieval models]{\label{fig:example:existing}
    \includegraphics[,scale=0.35]{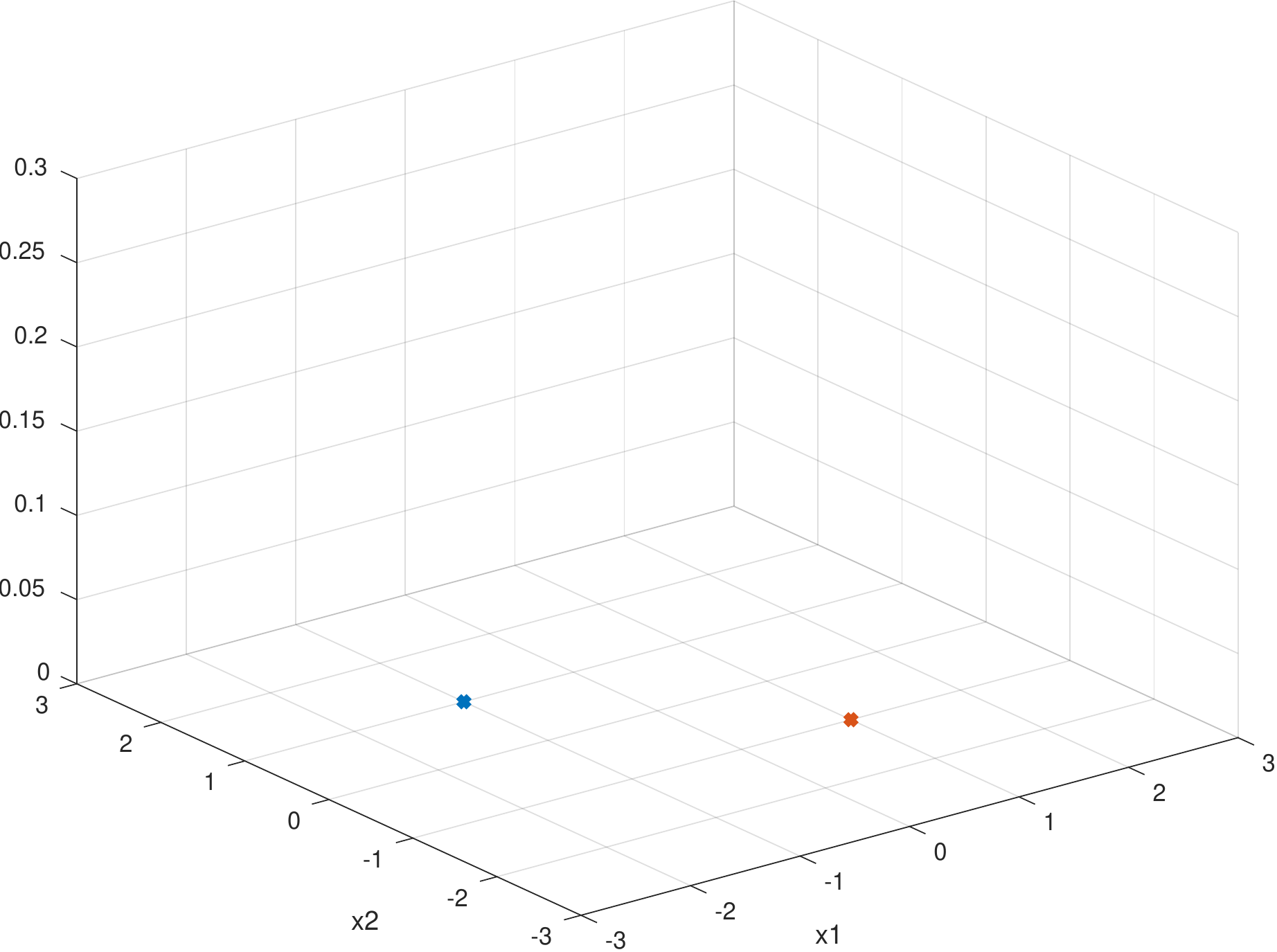}}
    \subfigure[Representation learning in \framework]{\label{fig:example:ours}        \includegraphics[trim="0cm 1.3cm 0cm 1cm",clip,scale=0.35]{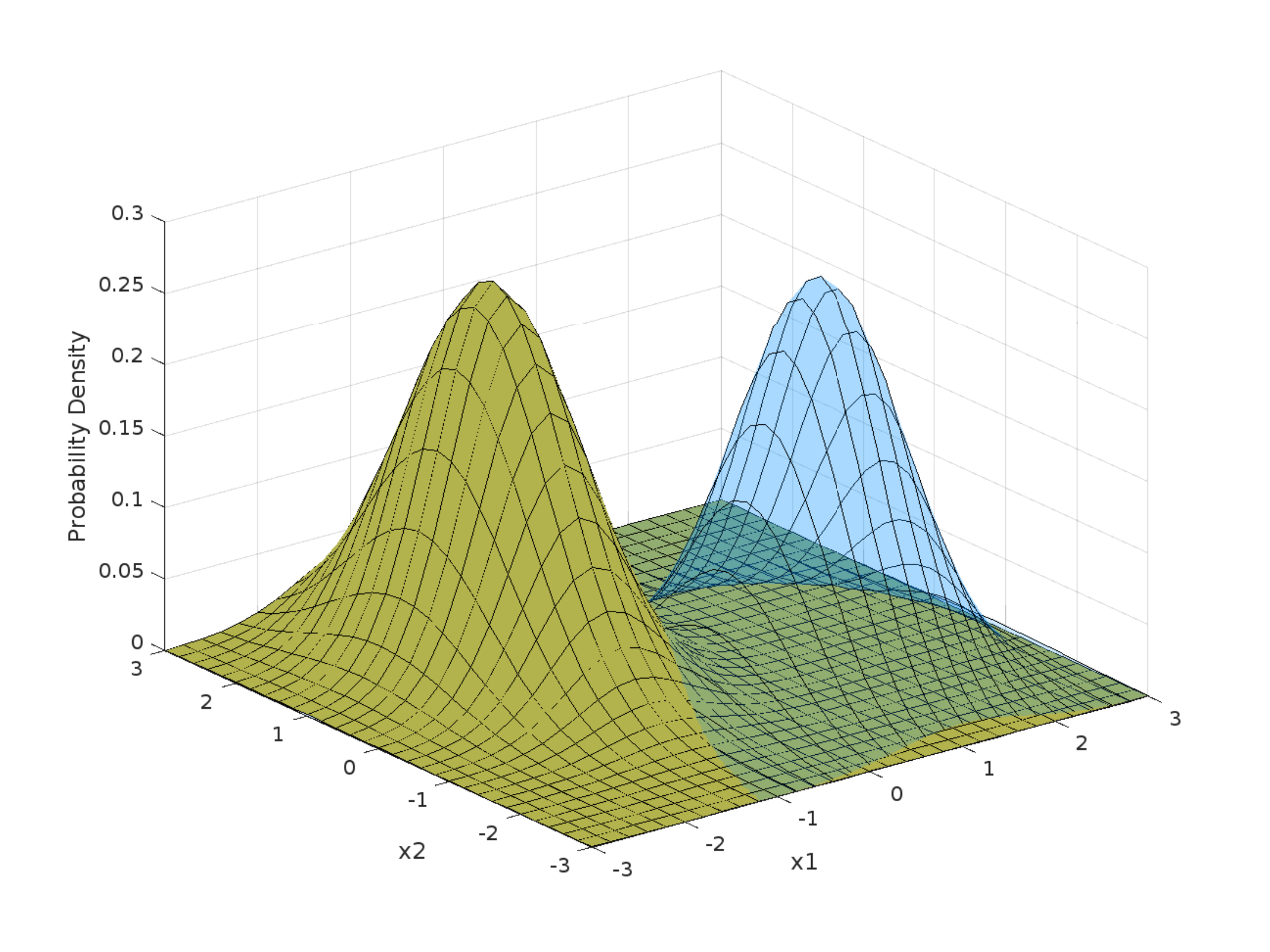}}
    \caption{Existing dense retrieval methods use a vector to represent any input. Figure \ref{fig:example:existing} demonstrates example representations they learn for two inputs (e.g., a query and a document). The proposed framework learns multivariate distributions to represent each input, which is depicted in Figure~\ref{fig:example:ours}.}
    \label{fig:example}
\end{figure*}

Therefore, we propose a generic framework that represents each query and document using a multivariate distribution, called the \framework framework. In other words, instead of representing queries and documents using $k$-dimensional vectors, we can assign a probability to each point in this $k$-dimensional space; the higher the probability, the higher the confidence that the model assigns to each point. For $k=2$, Figure~\ref{fig:example:existing} depicts the representation of a query and a document in existing single-vector dense retrieval models.\footnote{The third dimension is only used for consistent presentation. One can consider the probability of 1 for one point in the two-dimensional space and zero elsewhere.} On the other hand, Figure~\ref{fig:example:ours} demonstrates the representations that we envision for queries and documents.

To reduce the complexity of the model, we assume that the representations are multivariate normal distributions with a diagonal covariance matrix; meaning that the representation dimensions are orthogonal and independent. With this assumption, we learn two $k$-dimensional vectors for each query or document: a mean vector and a variance vector. In addition to uncertainty, such probabilistic modeling can implicitly represent breadth of information in queries and documents. For instance, a document that covers multiple topics and potentially satisfies a diverse set of information needs may be represented by a multivariate distribution with large variance values.

\framework uses negative multivariate Kullback-Leibler (KL) divergence between query and document representations to compute the relevance scores. We prove that the relevance scores can be computed efficiently by proposing solutions that can be implemented using existing approximate nearest neighbor search algorithms. We also demonstrate that one can simply implement the \framework framework using existing pre-trained large language models, such as BERT~\cite{BERT}. 

We show that an implementation of \framework that uses a single vector with 768 dimensions to represent multivariate representations for each query and document significantly outperforms existing single vector dense retrieval models on several standard text retrieval benchmarks. \framework also often outperforms ColBERTv2 \cite{Santhanam2021ColBERTv2EA}, a state-of-the-art multi vector dense retrieval model, while using significantly less storage and having significantly lower query latency. We further demonstrate that \framework also performs effectively in zero-shot settings when applied to unseen domains. Besides, we also demonstrate that the norm of variance vectors learned by \framework are a strong indicator of the retrieval effectiveness and can be used as a pre-retrieval query performance predictor.

We believe that \framework smooths the path towards developing more advanced probabilistic dense retrieval models and its applications can be extended to recommender systems, conversational systems, and a wide range of retrieval-enhanced machine learning models. 


\section{Related Work}

Variance of retrieval performance among different topics has been a long-standing research theme in the information retrieval community. For instance, TREC 2004 Robust Track organizers noted that solely optimizing the average metric aggregates (e.g., MAP) ``further improves the effectiveness of the already-effective topics, sometimes at the expense of the poor performers''~\cite{trec2004}. Moreover, identifying poorly performing topics is hard, and failure to do so leads to degraded user perception of the retrieval system as ``an individual user does not see the average
performance of the system, but only the effectiveness of the system on his or her requests''~\cite{trec2004}. These insights led the information retrieval community to consider \emph{query performance prediction}~\cite{Carmel:2010} -- a notion that certain signals can predict the performance of a search query. Such predictions can be helpful in guiding the retrieval system in taking further actions as needed for more difficult queries, e.g., suggesting alternative query reformulations~\cite{arabzadeh2021ms}.

A degree of query ambiguity with respect to the underlying corpus has been shown to be a valuable predictor of poor performance of search queries~\cite{Clarity}. Therefore, dealing with retrieval uncertainty has been proposed as a remedy. For instance, Collins-Thompson and Callan~\cite{collins+callan:2007} propose estimating query uncertainty by repeatedly fitting a Dirichlet distribution over bootstrap samples from the top-$k$ retrieved documents. They show that a Bayesian combination of multiple boostrap samples (which takes into account sample variance) leads to both significantly better retrieval metrics, and better retrieval robustness (less queries hurt by the query expansion methods). In a related vein, Zhu et al.~\cite{Zhu+al:2009} develop a risk-aware language model based on the Dirichlet distribution (as a conjugate prior to the multinomial distribution). They use the variance of the Dirichlet distribution for adjusting the risk in the final ranking score (i.e., revising the relevance estimates downwards in face of high variance). 

The idea of risk adjustment inspired by the financial investment literature was further developed by Wang and Zhu into the portfolio theory for information retrieval~\cite{Wang+Zhu:2009}. Portfolio theory generalizes the probability ranking principle (PRP) by considering both the uncertainty of relevance predictions and correlations between retrieved documents. It also demonstrates that one way to address uncertainty is via diversification~\cite{Clarke+al:2008}. The portfolio theory-based approach to retrieval has since been applied in several domains including recommendation~\cite{Shi+al:2012}, quantum-based information retrieval~\cite{Zuccon+al:2010}, and computational advertising~\cite{Zhang+al:2009}, among others. 

While, as this prior research shows, there has been an extensive exploration of risk and mean-variance trade-offs in the statistical language models for information retrieval, there has been so far much less discussion of these topics in the context of neural (\emph{aka} dense) models for retrieval. As a notable exception to this, \citet{Cohen+al:2021} recently proposed a Bayesian neural relevance model, where a posterior is approximated using Monte Carlo sampling based on drop-out~\cite{gal2016dropout}. A similar approach was proposed by \citet{penha2021calibration} in the context of conversational search. These approaches, which employ variational inference at training time, can only be applied for \emph{reranking}. In contrast, in this work we model uncertainty at the level of query and document representations, and demonstrate how such representations can be efficiently and effectively used for \emph{retrieval} using any of the existing approximate nearest neighbor methods.

Outside the realm of information retrieval research, various forms of representations that go beyond Euclidean vectors have been explored, including order embeddings~\cite{vendrov2015order}, hyperbolic embeddings~\cite{nickel2017poincare}, and probabilistic box embeddings~\cite{Vilnis+al:2018}, Such representations have been shown to be effective for various NLP tasks that involve modeling complex relationship or structures. Similar to our work, \citet{Vilnis+McCallum:2015} used Gaussian distributions for representation learning by proposing Gaussian embeddings for words. In this work, we focus on query and document representations in the retrieval setting.

Some prior work, as a way to achieve semantically richer representations, model queries and documents using a combination of multiple vectors~\cite{Kong+al:2022,Zhou+Devlin:2021,Santhanam2021ColBERTv2EA}. While such representations were shown to lead to better retrieval effectiveness, they do come at significant computational and storage costs. We demonstrate that our multivariate distribution representations are significantly more efficient than multi-vector ones, while attaining comparable or better performance on a wide range of collections.

\section{The \framework Framework}
\label{sec:framework}
Existing single vector dense retrieval models uses a $k$-dimensional latent vector to represent each query or each query token \cite{Xiong2021ApproximateNN,Hofsttter2020ImprovingEN,DPR,Zeng:2022:CLDRD}. We argue that these dense retrieval models can benefit from modeling uncertainty in representation learning. That means the model may produce a representation for a clear navigational query with high confidence, while it may have lower confidence in representing an ambiguous query. Same argument applies to the documents. However, the existing frameworks for dense retrieval do not model such confidence or uncertainty in representations. In this paper, we present \framework -- a generic framework for modeling uncertainty in representation learning for information retrieval. \framework models each query (or document) using a \emph{$k$-variate distribution} -- a group of $k$ continuous random variables using which we can compute the probability of any given vector in a $k$-dimensional space being a representation of the input query (or document). Formally, \framework encodes each query $q$ and each document $d$ as follows:
\begin{align}
\mathbf{Q} &= (Q_1, Q_2, \cdots, Q_k)^\intercal = \textsc{Encoder}_\textsc{q}(q) \nonumber \\
\mathbf{D} &= (D_1, D_2, \cdots, D_k)^\intercal = \textsc{Encoder}_\textsc{d}(d) 
\label{eq:encoders}
\end{align}
where $\textsc{Encoder}_\textsc{q}$ and $\textsc{Encoder}_\textsc{d}$ respectively denote query and document encoders. Each $Q_i$ and $D_i$ is a random variable; thus $\mathbf{Q}$ and $\mathbf{D}$ are $k$-variate distributions representing the query and the document. The superscript $\intercal$ denotes the transpose of the vector.

In this paper, we assume that $\mathbf{Q}$ and $\mathbf{D}$ both are $k$-variate normal distributions. The reasons for this assumption are: (1) we can define each $k$-variate normal distribution using a mean vector and a covariance matrix, (2) lower order distributions (i.e., any combination of the $k$ dimensions) and conditional distributions are also normal, which makes it easily extensible, (3) linear functions of multivariate normal distributions are also multivariate normal, leading to simple aggregation approaches. 
A $k$-variate normal distribution can be represented using a $k \times 1$ mean vector $\pmb{M} = (\mu_1, \mu_2, \cdots, \mu_k)^\intercal$ and a $k \times k$ covariance matrix $\pmb{\Sigma}$ as follows: $\mathcal{N}_k (\pmb{M}, \pmb{\Sigma})$. We compute the representations as $k$ \emph{independent} normal distributions, thus \textbf{the covariance matrix is diagonal}. Therefore, our representations are modeled as follows:
\begin{align}
    \mathcal{N}_k
    \begin{bmatrix}
    \begin{pmatrix}
    \mu_1\\
    \mu_2\\
    \vdots\\
    \mu_k
    \end{pmatrix}, &
    \begin{pmatrix}
    \sigma_1^2 & 0 & \dots & 0\\
    0 & \sigma_2^2 & \dots & 0\\
    \vdots & & \ddots & \\
    0 & 0 & \dots& \sigma_k^2
    \end{pmatrix}
    \end{bmatrix}
    \label{eq:kvarnormal}
\end{align}

With this formulation, we can re-write Equation~\eqref{eq:encoders} as follows:
\begin{align}
\mathbf{Q} &\sim \mathcal{N}_k(\pmb{M}_Q, \pmb{\Sigma}_Q), \qquad \pmb{M}_Q, \pmb{\Sigma}_Q = \textsc{Encoder}_\textsc{q}(q) \nonumber \\
\mathbf{D} &\sim \mathcal{N}_k(\pmb{M}_D, \pmb{\Sigma}_D), \qquad \pmb{M}_D, \pmb{\Sigma}_D = \textsc{Encoder}_\textsc{d}(d) 
\label{eq:encoders2}
\end{align}
where $\pmb{M}_Q = (\mu_{q1}, \mu_{q2}, \cdots, \mu_{qk})^\intercal$, $\pmb{\Sigma}_Q = (\sigma_{q1}^2, \sigma_{q2}^2, \cdots, \sigma_{qk}^2)^\intercal \times I_k$, $\pmb{M}_D = (\mu_{d1}, \mu_{d2}, \cdots, \mu_{dk})^\intercal$, and $\pmb{\Sigma}_D = (\sigma_{d1}^2, \sigma_{d2}^2, \cdots, \sigma_{dk}^2)^\intercal \times I_k$.
Therefore, it is safe to claim that \framework uses large language models to learn a $k$-dimensional mean vector and a $k$-dimensional variance vector for representing each input query and document. This representation for a query and a document  is plotted in Figure~\ref{fig:example:ours} ($k=2$ in the plot).

Using the flexible modeling offered by the \framework framework, we can compute the probability of any $k$ dimensional vector representing each query or document. In more detail, the probability of vector $\textbf{x} = (x_1, x_2, \cdots, x_k)^\intercal$ being generated from the $k$-variate normal distribution in Equation~\eqref{eq:kvarnormal} is equal to:
\begin{equation}
    p(\textbf{x}) = \frac{1}{(2\pi)^{\frac{k}{2}} \det(\pmb{\Sigma})^\frac{1}{2}} \exp{\left( -\frac{1}{2}(\textbf{x} - \pmb{M})^\intercal \pmb{\Sigma}^{-1} (\textbf{x} - \pmb{M}) \right)}
    \label{eq:pdf}
\end{equation}
where $\det(\cdot)$ denotes the determinant of the given matrix. This formulation enables us to compute the probability of any $k$-dimensional vector being a representation for each query and document. 

Once the queries and documents are represented, \framework computes the relevance score for a pair of query and document using the negative Kullback-Leibler divergence (negative KL divergence) between two $k$-variate distributions: $- \infdiv{\mathbf{Q}}{\mathbf{D}}$. The KL divergence can be computed as follows:
\begin{align}
    \infdiv{\mathbf{Q}}{\mathbf{D}} &= \mathbb{E}_{\mathbf{Q}} \left [\log \frac{\mathbf{Q}}{\mathbf{D}} \right] = \mathbb{E}_{\mathbf{Q}} \left [\log \mathbf{Q} - \log \mathbf{D} \right] \nonumber \\
    &= \frac{1}{2} \mathbb{E}_\mathbf{Q} \left [ -\log \det(\pmb{\Sigma}_Q) - (\textbf{x} - \pmb{M}_Q)^\intercal \pmb{\Sigma}_Q^{-1} (\textbf{x} - \pmb{M}_Q) \right. \nonumber \\
    &\quad\quad\quad\quad    \left. + \log \det(\pmb{\Sigma}_D) + (\textbf{x} - \pmb{M}_D)^\intercal \pmb{\Sigma}_D^{-1} (\textbf{x} - \pmb{M}_D) \right ] \nonumber \\
    &= \frac{1}{2} \log \frac{\det(\pmb{\Sigma}_D)}{\det(\pmb{\Sigma}_Q)} - \frac{1}{2} \mathbb{E}_\mathbf{Q} \left [ (\textbf{x} - \pmb{M}_Q)^\intercal \pmb{\Sigma}_Q^{-1} (\textbf{x} - \pmb{M}_Q) \right] \nonumber \\
    &\quad\quad\quad\quad\quad\quad\quad   +\frac{1}{2} \mathbb{E}_\mathbf{Q} \left[ (\textbf{x} - \pmb{M}_D)^\intercal \pmb{\Sigma}_D^{-1} (\textbf{x} - \pmb{M}_D) \right ] \label{eq:kldiv_1}
%
%
\end{align}

Since $(\textbf{x} - \pmb{M}_Q)^\intercal \pmb{\Sigma}_Q^{-1} (\textbf{x} - \pmb{M}_Q)$ is a real scalar (i.e., $\in \mathbb{R}$, it is equivalent to $\tr\{(\textbf{x} - \pmb{M}_Q)^\intercal \pmb{\Sigma}_Q^{-1} (\textbf{x} - \pmb{M}_Q) \}$, where $\tr\{\cdot\}$ denotes the trace of the given matrix. Since $\tr\{XY\} = \tr\{YX\}$ for any two matrices $X \in \mathbb{R}^{a \times b}$ and $Y \in \mathbb{R}^{b \times a}$, we have:
\begin{equation*}
    \tr\{(\textbf{x} - \pmb{M}_Q)^\intercal \pmb{\Sigma}_Q^{-1} (\textbf{x} - \pmb{M}_Q) \} = \tr\{(\textbf{x} - \pmb{M}_Q) (\textbf{x} - \pmb{M}_Q)^\intercal \pmb{\Sigma}_Q^{-1} \}
\end{equation*}

Therefore, since $\mathbb{E}[\tr\{X\}] = \tr\{\mathbb{E}[X]\}$ for any square matrix $X$, we can rewrite Equation~\eqref{eq:kldiv_1} as follows:
\begin{align}
    \infdiv{\mathbf{Q}}{\mathbf{D}} = \frac{1}{2} &\log \frac{\det(\pmb{\Sigma}_D)}{\det(\pmb{\Sigma}_Q)} \nonumber \\
    &  - \frac{1}{2} \tr \left\{\mathbb{E}_\mathbf{Q} \left [ (\textbf{x} - \pmb{M}_Q) (\textbf{x} - \pmb{M}_Q)^\intercal \pmb{\Sigma}_Q^{-1}  \right] \right\} \nonumber \\
    & + \frac{1}{2} \tr \left\{\mathbb{E}_\mathbf{Q} \left[  (\textbf{x} - \pmb{M}_D)^\intercal \pmb{\Sigma}_D^{-1} (\textbf{x} - \pmb{M}_D) \right ] \right\}
    \label{eq:kldiv_2}
\end{align}

Given the definition of the covariance matrix, we know that $\pmb{\Sigma}_Q = \mathbb{E}_\mathbf{Q} \left [ (\textbf{x} - \pmb{M}_Q) (\textbf{x} - \pmb{M}_Q)^\intercal \right]$. Therefore, we have:
\begin{align}
    \tr \{\mathbb{E}_\mathbf{Q} &\left [ (\textbf{x} - \pmb{M}_Q) (\textbf{x} - \pmb{M}_Q)^\intercal \pmb{\Sigma}_Q^{-1}  \right]  \} \nonumber \\
    &= \tr \left\{\mathbb{E}_\mathbf{Q} \left [ \pmb{\Sigma}_Q \pmb{\Sigma}_Q^{-1}  \right] \right \} = \tr \{I_k\} = k
    \label{eq:cond1}
\end{align}

In addition, since $Q$ is a multivariate normal distribution, for any matrix $A$ we have $\mathbb{E}_Q[\mathbf{x}^\intercal A \mathbf{x}] = \tr\{ A \pmb{\Sigma}_Q \} + \pmb{M}_Q^\intercal A \pmb{M}_Q$. This results in:
\begin{align}
    \tr \left\{\mathbb{E}_\mathbf{Q} \left [  (\textbf{x} - \pmb{M}_D)^\intercal \pmb{\Sigma}_D^{-1} (\textbf{x} - \pmb{M}_D) \right] \right \} = \nonumber \\
    \tr \{ \pmb{\Sigma}_D^{-1} \pmb{\Sigma}_Q \} + (\pmb{M}_Q - \pmb{M}_D)^\intercal  &\pmb{\Sigma}_D^{-1} (\pmb{M}_Q - \pmb{M}_D)
    \label{eq:cond2}
\end{align}

Using Equations \eqref{eq:cond1} and \eqref{eq:cond2}, we can rewrite Equation~\eqref{eq:kldiv_2} as follows:
\begin{align}
    \frac{1}{2} \left[\log \frac{\det(\pmb{\Sigma}_D)}{\det(\pmb{\Sigma}_Q)} - k + \tr \{ \pmb{\Sigma}_D^{-1} \pmb{\Sigma}_Q \} + (\pmb{M}_Q - \pmb{M}_D)^\intercal \pmb{\Sigma}_D^{-1} (\pmb{M}_Q - \pmb{M}_D) \right]
    \label{eq:kldiv_3}
\end{align}

This equation can be further simplified. Based on our earlier assumption that the covariance matrices are diagonal, then $\det(\pmb{\Sigma}_D) = \prod_{i=1}^{k}{\sigma_{di}^2}$. In addition, since we are using KL divergence to rank documents, constant values (e.g., $k$) or document independent values (e.g., $\log \det(\pmb{\Sigma}_Q)$) do not impact document ordering. Therefore, there can be omitted and we can use the following equation to rank the documents using negative multivariate KL-divergence:
\begin{align}
    \text{score}(q, d) &= - \infdiv{\mathbf{Q}}{\mathbf{D}} \nonumber \\
    &=^{\text{rank}} -\frac{1}{2} \left[ \sum_{i=1}^{k}{\log \sigma_{di}^2} + \frac{\prod_{i=1}^{k}{\sigma_{qi}^2}}{\prod_{i=1}^{k}{\sigma_{di}^2}} + \sum_{i=1}^{k}{\frac{(\mu_{qi} - \mu_{di})^2}{\sigma_{di}^2}}     \right] 
    \label{eq:kldiv_final}
\end{align}

In Section~\ref{sec:ann}, we explain how to efficiently compute this scoring function using approximate nearest neighbor methods.

\section{\framework Implementation}
\label{sec:implementation}
In this section, we first describe our network architecture for implementing the the query and document encoders $\textsc{Encoder}_\textsc{q}$ and $\textsc{Encoder}_\textsc{d}$ (see Equation~\eqref{eq:encoders2}). Next, we explain our optimization approach for training the models. 

\subsection{Encoder Architecture}
Pretrained large language models (LLMs) have demonstrated promising results in various information retrieval tasks \cite{Nogueira2019PassageRW,Zeng:2022:CLDRD,Hofsttter2020ImprovingEN,trec2020overview}. Therefore, we decide to adapt existing pretrained LLMs to learn a $k$-variate normal distribution for each given input. As described above, each $k$-variate normal distribution can be modeled using a $k$-dimensional mean vector and a $k$-dimensional variance vector. We use two special tokens as the input of pretrained LLMs to obtain these two vectors. For example, we convert an input query `\texttt{neural information retrieval}' to `\texttt{[CLS] [VAR] neural information retrieval [SEP]}' and feed it to BERT-base \cite{BERT}. Let $\vec{q}_\texttt{[CLS]} \in \mathbb{R}^{1 \times 768}$ and $\vec{q}_\texttt{[VAR]} \in \mathbb{R}^{1 \times 768}$ respectively denote the representations produced by BERT for the first two tokens \texttt{[CLS]} and \texttt{[VAR]}. We obtain the mean and variance vectors for query $q$ using two separate dense projection layers on $\vec{q}_\texttt{[CLS]}$ and $\vec{q}_\texttt{[VAR]}$, as follows:
\begin{align}
    \pmb{M}_Q &= \vec{q}_\texttt{[CLS]} W_M \nonumber \\
    \pmb{\Sigma}_Q &= \frac{1}{\beta} \log(1 + \exp (\beta . \vec{q}_\texttt{[VAR]} W_\Sigma)) . I_k 
    \label{eq:encoder3}
\end{align}
where $W_M \in \mathbb{R}^{768 \times k}$ and $W_\Sigma \in \mathbb{R}^{768 \times k}$ are the projection layer parameters. To compute the diagonal covariance matrix, we use the softplus function (i.e., $\frac{1}{\beta}\log(1 + \exp (\beta . x))$) for the following reasons: (1) it is continuous and differentiable, thus it can be used in gradient descent-based optimization, (2) softplus ensures that variance values are always positive, (3) zero is its lower bound ($\lim_{x \rightarrow -\infty}\frac{1}{\beta}\log(1 + \exp (\beta . x)) = 0$), yet it is never equal to zero, thus it does not cause numeric instability in KL-divergence calculation (see Equation~\eqref{eq:kldiv_final}), and (4) for large $x$ values, it can be approximated using a linear function, i.e., $\lim_{x \rightarrow \infty}\frac{1}{\beta}\log(1 + \exp (\beta . x)) = x$, ensuring numerical stability for large input values. To better demonstrate its properties, Figure~\ref{fig:softplus} in our experiments plots softplus for various values of $\beta$ -- a hyper-parameter that specifies the softplus formation. The $k \times k$ identity matrix $I_k$ in Equation~\eqref{eq:encoder3} is used to convert the variance vector to a diagonal covariance matrix. 

Note that \framework does not explicitly compute variance, instead learns representations for the \texttt{[VAR]} token such that it minimizes the loss function based on negative multivariate KL divergence scoring. Therefore, the model implicitly learns how to represent latent variance vectors.

The mean vector and covariance matrices for document representations are also computed similarly. In our experiments, all parameters (including parameters in BERT and the dense projection layers) are updated and shared between the query and document encoders (i.e., $\textsc{Encoder}_\textsc{q}$ and $\textsc{Encoder}_\textsc{d}$).

\subsection{Model Training}
Recent research has suggested that dense retrieval models can significantly benefit from knowledge distillation \cite{Hofsttter2021EfficientlyTA,Santhanam2021ColBERTv2EA,Qu2021RocketQAAO}. Following these models, we use a BERT-based cross-encoder re-ranking model as the teacher model. Let $D_{q}$ be a set of documents selected for query $q$ for knowledge distillation. We use the following listwise loss function for each query $q$ as follows:
\begin{equation}
    \sum_{d, d' \in D_{q}}  \mathbbm{1}\{y^T_q(d) > y^T_q(d')\} |\frac{1}{\pi_q(d)} - \frac{1}{\pi_q(d')}| \log ( 1+ e^{y^S_q(d') - y^S_q(d)})
\end{equation}
where $\pi_q(d)$ denotes the rank of document $d$ in the result list produced by the student dense retrieval model, and $y^T_q(d)$ and $y^S_q(d)$ respectively denote the scores produced by the teacher and the student models for the pair of query $q$ and document $d$. This knowledge distillation listwise loss function is inspired by LambdaRank \cite{Burges2010FromRT} and is also used by \citet{Zeng:2022:CLDRD} for dense retrieval distillation.

For each query $q$, the document set $D_q$ is constructed based on the following steps:
\begin{itemize}
    \item $D_q$ includes all positive documents from the relevance judgments (i.e., qrel).
    \item $D_q$ includes $m_{\text{BM25}} \in \mathbb{R}$ documents from the top 100 documents retrieved by BM25. 
    \item $D_q$ includes $m_{\text{hard}} \in \mathbb{R}$ documents from the top 100 documents retrieved by student model (i.e., negative sampling using the model itself every 5000 steps).
\end{itemize}

In addition, we take advantage of the other passages in the batch as in-batch negatives. Although in-batch negatives resemble randomly sampled negatives that can be distinguished easily from other documents, it is efficient since passage representations can be reused within the batch \cite{DPR}.

\subsection{Efficient Retrieval}
\label{sec:ann}
Existing dense retrieval models use approximate nearest neighbor (ANN) approaches for efficient retrieval. However, using ANN algorithms in the proposed \framework framework is not trivial. The reason is that \framework uses the negative $k$-variate KL divergence formulation presented in Equation~\eqref{eq:kldiv_final} to compute relevance scores.  This is while existing ANN algorithms only support simple similarity functions such as dot product, cosine similarity, or negative Euclidean distance. 
To address this issue, we convert Equation~\eqref{eq:kldiv_final} to a dot product formulation. Let us expand the last term in Equation~\eqref{eq:kldiv_final}:\footnote{We drop multiplication to $\frac{1}{2}$ as it does not impact document ordering.}
\begin{equation}
    - \left[ \underbrace{\sum_{i=1}^{k}{\log \sigma_{di}^2}}_{\text{doc prior}} + \frac{\prod_{i=1}^{k}{\sigma_{qi}^2}}{\prod_{i=1}^{k}{\sigma_{di}^2}} + \sum_{i=1}^{k}{\frac{\mu_{qi}^2}{\sigma_{di}^2}} + \underbrace{\sum_{i=1}^{k}{\frac{\mu_{di}^2}{\sigma_{di}^2}}}_{\text{doc prior}} - \sum_{i=1}^{k}{\frac{2\mu_{di}\mu_{qi}}{\sigma_{di}^2}}    \right]
    \label{eq:kldiv_efficient}
\end{equation}

The first and the fourth terms in Equation~\eqref{eq:kldiv_efficient} are document priors, thus they are query independent and can be pre-computed. Therefore, let $\gamma_d = - \sum_{i=1}^{k}{(\log \sigma_{di}^2 +\frac{\mu_{di}^2}{\sigma_{di}^2})}$ denote the document prior score. Therefore, the scoring function in Equation~\eqref{eq:kldiv_final} can be formulated as the dot product of the following two vectors:
\begin{align}
 \vec{q} &= \left[1, \Pi_q, \mu_{q1}^2, \mu_{q2}^2, \cdots, \mu_{qk}^2, \mu_{q1}, \mu_{q2}, \cdots, \mu_{qk} \right]  \nonumber \\
 \vec{d} &= \left[\gamma_d, \frac{-1}{\Pi_d}, \frac{-1}{\sigma_{d1}^2}, \frac{-1}{\sigma_{d2}^2}, \cdots, \frac{-1}{\sigma_{dk}^2}, \frac{2\mu_{d1}}{\sigma_{d1}^{2}}, \frac{2\mu_{d2}}{\sigma_{d2}^{2}}, \cdots, \frac{2\mu_{dk}}{\sigma_{dk}^{2}}\right] 
\end{align}
where $\Pi_q = \prod_{i=1}^{k}{\sigma_{qi}^2}$ and $\Pi_d = \prod_{i=1}^{k}{\sigma_{di}^2}$ are pre-computed scalars. 
The dot product of $\vec{q} \in \mathbb{R}^{1 \times (2k+2)}$ and $\vec{d} \in \mathbb{R}^{1 \times (2k+2)}$ is equal to the retrieval score formulated in Equation~\eqref{eq:kldiv_final}. More importantly, $\vec{q}$ is document independent and $\vec{d}$ is query independent. Therefore, we can use existing approximate nearest neighbor algorithms, such as HNSW \cite{HNSW}, and existing tools, such as FAISS \cite{FAISS}, to index all $\vec{d}$ vectors and conduct efficient retrieval for any query vector $\vec{q}$.

\section{Discussion}
\label{sec:discussion}

In this section, we attempt to shed some light on the behavior of retrieval using \framework, by providing theoretical answers to the following questions.

\medskip

\begin{quote}
    Q1. How does \framework rank two documents with identical covariance matrices?
\end{quote}
Let $d$ and $d'$ be two documents, represented by the mean vectors $\pmb{M}_D$ and $\pmb{M}_{D'}$ and identical covariance matrix $\pmb{\Sigma}_D = \pmb{\Sigma}_{D'}$. Therefore, given Equation~\eqref{eq:kldiv_final} we have:
\begin{align*}
    \text{score}(q, d) - \text{score}(q, d') =^{\text{rank}}  \sum_{i=1}^{k}{\left[(\mu_{qi} - \mu_{d'i})^2 - (\mu_{qi} - \mu_{di})^2 \right]} 
\end{align*}

This shows that in case of identical covariance matrices, \framework assigns a higher relevance score to the document whose mean vector is closest to the query mean vector with respect to \emph{Euclidean distance}.

A remark of this finding is that if the covariance matrix is constant for all documents (i.e., if we ignore uncertainty), \framework can be reduced to existing dense retrieval formulation, where negative Euclidean distance is used to measure vector similarity. Therefore, \framework is a generalized form of this dense retrieval formulation.

\medskip

\begin{quote}
    Q2. Popular dense retrieval models use inner product to compute the similarity between query and document vectors. What happens if we use inner product in \framework?
\end{quote}

\begin{table*}[t]
    \centering
    \caption{Characteristics and statistics of the datasets in our experiments.}
    \begin{tabular}{llccc}\toprule
        \textbf{Dataset} & \textbf{Domain} & \textbf{\# queries} & \textbf{\# documents} & \textbf{avg doc length} \\\midrule
        MS MARCO DEV & Miscellaneous & 6,980 & 8,841,823 & 56 \\
        TREC DL '19 & Miscellaneous & 43 & 8,841,823 & 56 \\
        TREC DL '20 & Miscellaneous & 54 & 8,841,823 & 56  \\\midrule
        SciFact & Scientific fact retrieval & 300 & 5,183 & 214  \\
        FiQA & Financial answer retrieval & 648 & 57,638 & 132  \\
        TREC COVID & Bio-medical retrieval for Covid-19& 50 & 171,332 & 161  \\
        CQADupStack & Duplicate question retrieval & 13,145 & 457,199 & 129 \\
        \bottomrule
    \end{tabular}
    \label{tab:data}
\end{table*}

Inner product or dot product cannot be defined for multivariate distributions, however, one can take several samples from the query and document distributions and compute their dot product similarity. Since the query distribution $\mathbf{Q} \sim \mathcal{N}_k(\pmb{M}_Q, \pmb{\Sigma}_Q)$ and the document distribution $\mathbf{D} \sim \mathcal{N}_k(\pmb{M}_D, \pmb{\Sigma}_D)$ are independent, the expected value of their product is:
$$\mathbb{E}[\mathbf{Q} \cdot \mathbf{D}] = \mathbb{E}[\mathbf{Q}] \cdot \mathbb{E}[\mathbf{D}] = \pmb{M}_Q \cdot \pmb{M}_D$$

That means, in expectation, the dot product of samples from multivariate distributions will be equivalent to the dot product of their mean vectors. Therefore, with this formulation (i.e., using expected dot product instead of negative KL divergence) the results produced by \framework will be equivalent to the existing dense retrieval models and representation uncertainties are not considered.

\medskip

\begin{quote}
    Q3. Negative KL divergence has been used in the language modeling framework of information retrieval \cite{Lafferty:2001}. How is it connected with the proposed \framework framework?
\end{quote}

\citet{Lafferty:2001} extended the query likelihood retrieval model of \citet{Ponte:1998} by computing negative KL divergence between unigram query and document language models. Similarly, \framework uses negative KL divergence to compute relevance scores, however, there are several fundamental differences. First, \citet{Lafferty:2001} compute the distributions based on term occurrences in queries and documents through maximum likelihood estimation, while \framework learns \emph{latent} distributions based on the contextual representations learned from queries and documents. Second, \citet{Lafferty:2001} use univariate distributions for queries and documents, while \framework uses high-dimensional multivariate distributions. 





    


\section{Experiments}
To evaluate the impact of multivariate representation learning, we first run experiments on standard passage retrieval collections from MS MARCO and TREC Deep Learning Tracks. We also study the parameter sensitivity of the model in this task. We further demonstrate the ability of multivariate representations to better model distribution shift when applied to zero-shot retrieval settings, i.e., retrieval on a target collection that is significantly different from the training set. Our experiments also shows that the norm of learned variance vectors is correlated with the retrieval performance of the model.

\subsection{Datasets}
\label{sec:data}
In this section, we introduce our training set and evaluation sets whose characteristics and statistics are reported in Table~\ref{tab:data}.

\paragraph{\textbf{Training Set.}} We train our ranking model on the MS MARCO passage retrieval training set. The MS MARCO collection \cite{MSMARCO} contains approximately 8.8M passages and its training set includes 503K unique queries. The MS MARCO training set was originally constructed for a machine reading comprehension tasks, thus it did not follow the standard IR annotation guidelines (e.g., pooling). The training set contains an average of 1.1 relevant passage per query, even though there exist several relevant documents that are left adjudged. This is one of the reasons that knowledge distillation help dense retrieval models learn more robust representations. 

\paragraph{\textbf{Passage Retrieval Evaluation Sets.}} We evaluate our models on three query sets for the passage retrieval task. They all use the MS MARCO passage collection. These evaluation query sets are: (1) \textit{MS MARCO DEV}: the standard development set of MS MARCO passage retrieval task that consists of 6980 queries with incomplete relevance annotations (similar to the training set), (2) \textit{TREC-DL'19}: passage retrieval query set used in the first iteration
of TREC Deep Learning Track in 2019 \cite{trec2019overview} which includes 43 queries, and (3) \textit{TREC-DL'20}: the passage retrieval query set of TREC Deep Learning Track 2020 \cite{trec2020overview} with 54 queries. Relevance annotation for TREC DL tracks was curated using standard pooling techniques. Therefore, we can consider them as datasets with complete relevance annotations.

\begin{table*}[t]
     \centering
     \caption{The passage retrieval results obtained by the proposed approach and the baselines. The highest value in each column is bold-faced. The superscript $^*$ denotes statistically significant improvements compared to all the baselines based on two-tailed paired t-test with Bonferroni correction at the 95\% confidence level. ``-'' denotes the results that are not applicable or available.}
    \begin{tabular}{llllcccccc}
    \toprule
    \multirow{2}{*}{\textbf{Model}} &
     &
    \multirow{2}{*}{\textbf{Encoder}} &
    \multirow{2}{*}{\textbf{\#params}} & 
    \multicolumn{2}{c}{\textbf{MS MARCO DEV}} & 
       \multicolumn{2}{c}{\textbf{TREC-DL'19}}&
       \multicolumn{2}{c}{\textbf{TREC-DL'20}}\\
       
    & &  &   &  \textbf{MRR@10}  & \textbf{MAP}  & \textbf{NDCG@10} & \textbf{MAP} & \textbf{NDCG@10} & \textbf{MAP} \\\midrule
       \multicolumn{3}{l}{\textbf{Single Vector Dense Retrieval Models}} \\
        ANCE \cite{Xiong2021ApproximateNN}  &  &  BERT-Base & 110M & 0.330 & 0.336 & 0.648 & 0.371 & 0.646 & 0.408 \\
        ADORE \cite{Zhan2021OptimizingDR} &    & BERT-Base & 110M & 0.347  & 0.352 & 0.683 & 0.419 & 0.666 & 0.442 \\ 
         RocketQA \cite{Qu2021RocketQAAO} &  &   ERNIE-Base & 110M  & 0.370 & - & - & - & - & -\\ 
       Contriever-FT \cite{contriever} & & BERT-Base & 110M & - & - & 0.621 & - & 0.632 & - \\
        TCT-ColBERT \cite{Lin2020DistillingDR} &   &  BERT-Base & 110M  & 0.335 & 0.342 & 0.670 & 0.391 & 0.668 & 0.430 \\ 
        Margin-MSE \cite{Hofsttter2020ImprovingEN} &    &  DistilBERT  & 66M & 0.325 & 0.331 & 0.699 & 0.405 & 0.645 & 0.416 \\
        TAS-B \cite{Hofsttter2021EfficientlyTA} &  &  DistilBERT & 66M  &  0.344  & 0.351 & 0.717 & 0.447 & 0.685 & 0.455 \\ 
        
       
       CLDRD \cite{Zeng:2022:CLDRD} &  &   DistilBERT & 66M  & 0.382 & 0.386  & 0.725 & 0.453 & 0.687 & 0.465 \\
       \textbf{\framework (ours)} & & DistilBERT & 66M  & \textbf{0.393$^*$} & \textbf{0.402$^*$}  & \textbf{0.738} & \textbf{0.472$^*$} & 0.701$^*$ & \textbf{0.479$^*$} \\ \midrule
        \multicolumn{6}{l}{\textbf{Some Sparse Retrieval Models (For Reference)}} \\
        BM25 \cite{Robertson1995OkapiBM25} &   & - & -  & 0.187 & 0.196 & 0.497 & 0.290 & 0.487 & 0.288 \\ 
       DeepCT \cite{Dai2019ContextAwareST} &   &  BERT-Base & 110M  & 0.243 & 0.250 & 0.550 & 0.341 & 0.556 & 0.343 \\ 
        docT5query \cite{Nogueira2019FromDT}  &   &  T5-Base & 220M  & 0.272 & 0.281 & 0.642 & 0.403 & 0.619 & 0.407 \\\\
        \multicolumn{6}{l}{\textbf{Multi Vector Dense Retrieval Model (For Reference)}} \\
        ColBERTv2 \cite{Santhanam2021ColBERTv2EA} &  & DistilBERT & 66M & 0.384 & 0.389 & 0.733 & 0.464 & \textbf{0.712} & 0.473 \\ 
       \bottomrule
    \end{tabular}
    \label{tab:results}
\end{table*}

\paragraph{\textbf{Zero-Shot Passage Retrieval Evaluation Sets.}}
To demonstrate the generalization of retrieval models to different domains, we perform a zero-shot passage retrieval experiment (i.e., the models are trained on the MS MARCO training set). To do so, we use four domains which diverse properties. (1) \textit{SciFact} \cite{scifact}: a dataset for scientific fact retrieval with 300 queries, (2) \textit{FiQA} \cite{fiqa}: a passage retrieval dataset for natural language questions in the financial domain with 648 queries, (3) \textit{TREC COVID} \cite{trec-covid}: a task of retrieving abstracts of bio-medical articles in response to 50 queries related to the Covid-19 pandemic, and (4) \textit{CQADupStack} \cite{cqadupstack}: the task of duplicated question retrieval on 12 diverse StackExchange websites with 13,145 test queries. To be consistent with the literature, we used the BEIR \cite{BEIR} version of all these collections.

\subsection{Experimental Setup}
We implemented and trained our models using TensorFlow. The network parameters were optimized with Adam \cite{AdamOptimizer} with  linear scheduling with the warmup of $4000$ steps. In our experiments, the learning rate was selected from $[\num{1e-6}, \num{1e-5}]$ with a step size of $\num{1e-6}$. The batch size was set to $512$. The parameter $\beta$ was selected from $[0.5, 1, 2.5, 5, 7.5, 10]$. {To have a fair comparison with the baselines that often use 768 dimensions for representing queries and documents using BERT, we set the parameter $k$ (i.e., the number of random variables in our multivariate normal distributions) to $\mathbf{\frac{768}{2}-1=381}$ (see Section \ref{sec:ann} for more information).} In our experiments, we use the DistilBERT \cite{DistilBERT} with the pre-trained checkpoint made available from TAS-B \cite{Hofsttter2021EfficientlyTA} as the initialization. As the re-ranking teacher model, we use a BERT cross-encoder, similar to that of \citet{Nogueira2019PassageRW}. Hyper-parameter selection and early stopping was conducted based on the performance in terms of MRR on the MS MARCO validation set.

\subsection{Evaluation Metrics}
We use appropriate metrics for each evaluation set based on their properties. For MS MARCO Dev, we use MRR@10 which is the standard metric for this dataset, and we followed TREC Deep Learning Track's recommendation on using NDCG@10 \cite{NDCG} as the evaluation metrics. To complement our result analysis, we also use  mean average precision of the top 1000 retrieved documents (MAP), which is a common recall-oriented metric. For zero-shot evaluation, we follow BEIR's recommendation and use NDCG@10 to be consistent with the literature \cite{BEIR}.
The two-tailed paired t-test with Bonferroni correction is used to identify statistically significant performance differences ($p\_value < 0.05$).

\subsection{Experimental Results}

\paragraph{\textbf{Baselines.}} We also compare against the following state-of-the-art dense retrieval models with single vector representations:
\begin{itemize}[leftmargin=*]
    \item ANCE \cite{Xiong2021ApproximateNN} and ADORE \cite{Zhan2021OptimizingDR}: two effective dense retrieval models based on BERT-Base \cite{BERT} that use the model itself to mine hard negative documents.
    \item RocketQA \cite{Qu2021RocketQAAO}, Margin-MSE \cite{Hofsttter2020ImprovingEN}, and TAS-B \cite{Hofsttter2021EfficientlyTA}: effective dense retrieval models that use knowledge distillation from a BERT reranking model (a cross-encoder) in addition to various techniques for negative sampling.
    \item Contriever-FT \cite{contriever}: a single vector dense retrieval model that is pre-trained for retrieval tasks and then fine-tuned on MS MARCO. This model has shown effective performance on out-of-distribution target domain datasets. 
    \item TCT-ColBERT \cite{Lin2020DistillingDR}: a single vector dense retrieval model that is trained through knowledge distillation where a multi vector dense retrieval model (i.e., ColBERT \cite{Khattab2020ColBERTEA}) is used as the teacher model.
    \item CLDRD \cite{Zeng:2022:CLDRD}: the state-of-the-art single vector dense retrieval model that uses knowledge distillation from a reranking teacher model through gradual increase of training data difficulty (curriculum learning).
\end{itemize}

Even though \framework is a single vector dense retrieval model, as a point of reference, we use a state-of-the-art dense retrieval model with multiple vectors (i.e., ColBERTv2 \cite{Santhanam2021ColBERTv2EA}). For demonstrating a fair comparison, all baselines are trained and tuned in the same way as the proposed approach. 

We also compare our model against the following baselines that use inverted index for computing relevance scores (sometimes called sparse retrieval models): 
\begin{itemize}[leftmargin=*]
    \item BM25 \cite{Robertson1995OkapiBM25}: a simple yet strong term matching model for document retrieval that computes relevance scores based on term frequency in each document, document length, and inverse document frequency in the collection. We use the Galago search engine \cite{galago} to compute BM25 scores and tuned BM25 parameters using the training set. 
    \item DeepCT \cite{Dai2019ContextAwareST}: an approach that uses BERT to compute a weight for each word in the vocabulary for each document and query. Then words with highest weights are then selected and added to the inverted index with their weights. This approach can be seen as a contextual bag-of-words query and document expansion approach. 
    \item docT5query \cite{Nogueira2019FromDT}: a sequence-to-sequence model based on T5 \cite{T5} that is trained on MS MARCO to generate queries from any relevance passage. The documents are then expanded using the generated queries. 
\end{itemize}

\begin{table}[t]
    \centering
    \caption{A comparison of storage requirement and query latency between single vector and multi vector dense retrieval models with DistilBERT encoders on MS MARCO collection with 8.8 million passages. We ran this experiment on a machine with 16 Core i7-4790 CPU @ 3.60GHz.}
    \begin{tabular}{lcc}\toprule
         & \textbf{storage requirement} & \textbf{query latency} \\\midrule
        Single vector DR & 26GB & 89 ms / query \\
        Multi vector DR & 192GB & 438 ms / query \\\bottomrule
    \end{tabular}
    \label{tab:efficiency}
\end{table}

\paragraph{\textbf{The Passage Retrieval Results.}} The passage retrieval results are presented in Table~\ref{tab:results}. According to the table, all dense retrieval models perform substantially better than BM25 and DeepCT, demonstrating the effectiveness of such approaches for in-domain passage retrieval tasks. We observe that the approaches that use knowledge distillation (i.e., every dense retrieval model, except for ANCE, ADORE, and Contriever-FT) generally perform better than others. The recent CLDRD model shows the strongest retrieval results among all single vector dense retrieval models. The multi vector dense retrieval approach (ColBERTv2) outperforms all single vector dense retrieval baselines. Note that ColBERTv2 stores a vector for each token in the documents and thus it requires significantly larger storage for storing the ANN index and also suffers from substantially higher query latency (see Table~\ref{tab:efficiency} for more information). We show that \framework outperforms all baselines in terms of all the evaluation metrics used in the study. The improvements compared to all baselines are statistically significant, except for NDCG@10 in TREC-DL'19; the $p\_value$ (corrected using Bonferroni correction) for \framework versus CLDRD in this case was $0.07381$. Note that this dataset only contains 43 queries and significance tests are impacted by sampled size. \framework performs significantly better than any other baseline in this case. 

\begin{table}[t]
    \centering
    \caption{Sensitivity of \framework's retrieval performance to different values of $\beta$.}
    \resizebox{\linewidth}{!}{
    \begin{tabular}{lcccccc}\toprule
        & \multicolumn{2}{c}{\textbf{MS MARCO DEV}} & 
       \multicolumn{2}{c}{\textbf{TREC-DL'19}}&
       \multicolumn{2}{c}{\textbf{TREC-DL'20}}\\
        & \textbf{MRR@10}  & \textbf{MAP}  & \textbf{NDCG@10} & \textbf{MAP} & \textbf{NDCG@10} & \textbf{MAP} \\ \midrule
        $\beta = 0.1$ & 0.385 & 0.384 & 0.723 & 0.448 & 0.693 & 0.466 \\
        $\beta = 0.25$ & 0.399 & 0.415 & 0.743 & 0.468 & 0.704 & 0.478 \\
        $\beta = 0.5$ & 0.403 & 0.408 & 0.742 & 0.481 & 0.703 & 0.486 \\
        $\beta = 1$ & 0.403 & 0.412  & 0.748 & 0.480 & 0.711 & 0.489 \\
        $\beta = 5$ & 0.405 & 0.421 & 0.749 & 0.484 & 0.716 & 0.489 \\
        $\beta = 10$ & 0.402 & 0.421 & 0.758 & 0.489 & 0.701 & 0.483 \\
        \bottomrule
    \end{tabular}
    }
    \label{tab:sensitivity_beta}
\end{table}

\begin{figure}[t]
    \centering
    \includegraphics[width=.8\linewidth]{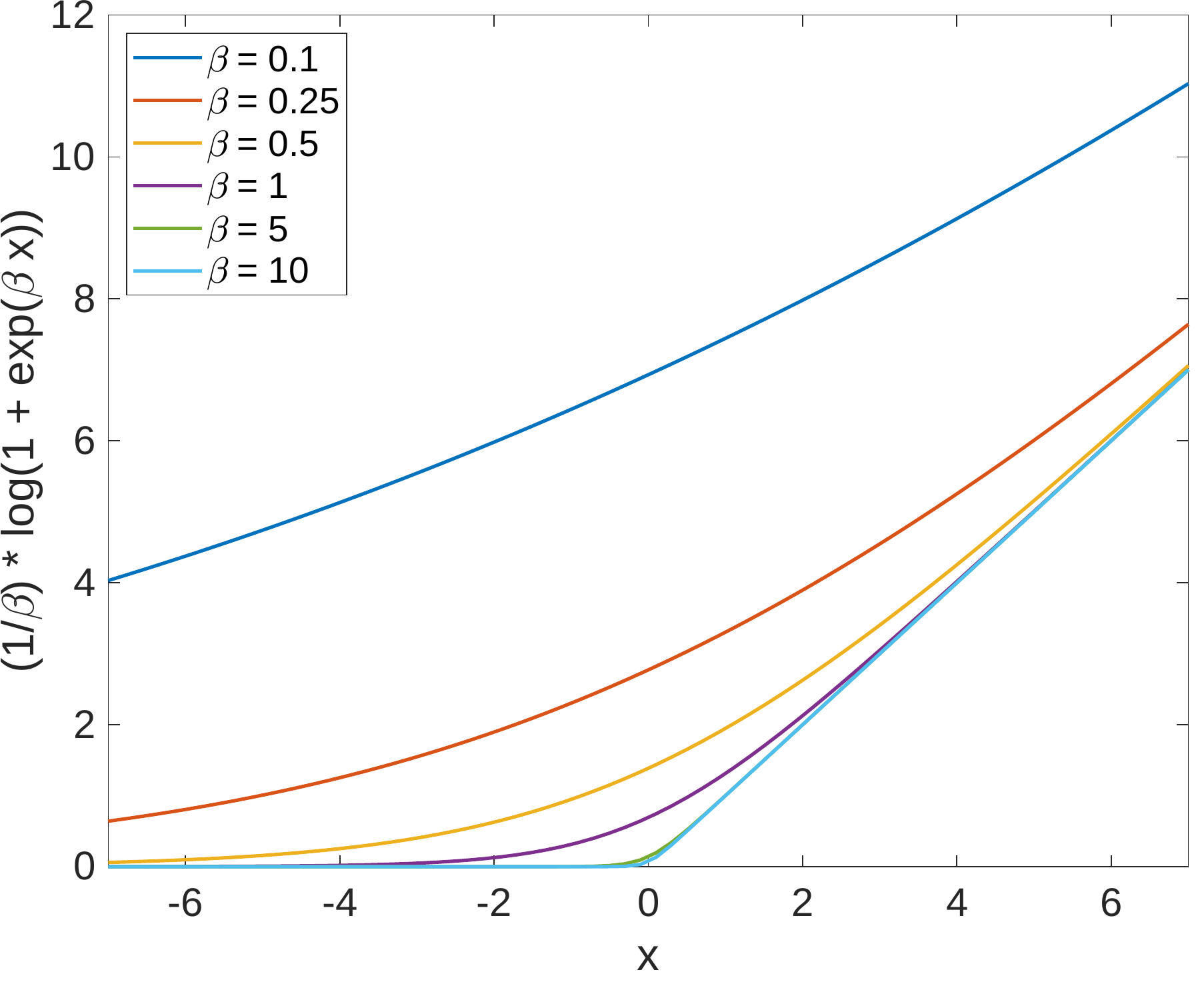}
    \caption{The softplus curve that is used to compute the variance vector for different values of $\beta$. Softplus is a monotonic and increasing function with a lower bound of zero. It's value for large $x$ values can be approximated using the linear function $y=x$ for numeric stability.}
    \label{fig:softplus}
\end{figure}

\paragraph{\textbf{Parameter Sensitivity Analysis.}} To measure the sensitivity of \framework's performance to the value of $\beta$, we change $\beta$ from 0.1 to 10 and report the results in Table~\ref{tab:sensitivity_beta}. To get a sense of the impact of these values, please see Figure~\ref{fig:softplus}. The results show that the model is not sensitive to the value of $\beta$ unless it is smaller than or equal to $\leq 0.25$. Therefore, for a $\beta$ value of around 1 or larger, the model shows a robust and strong performance.

\begin{table}[t]
    \centering
    \caption{The zero-shot retrieval results obtained by the proposed approach and the baselines, in terms of NDCG@10. The highest value in each column is bold-faced. The superscript $^*$ denotes statistically significant improvements compared to all the baselines based on two-tailed paired t-test with Bonferroni correction at the 95\% confidence level.}
    \begin{tabular}{lcccc}\toprule
       \multirow{2}{*}{\textbf{Model}} & \multirow{2}{*}{\textbf{SciFact}} & \multirow{2}{*}{\textbf{FiQA}} & \textbf{TREC} & \textbf{CQA} \\
         &  &  & \textbf{COVID} & \textbf{DupStack} \\\midrule
        \multicolumn{3}{l}{\textbf{Single Vector DR Models}} \\
        ANCE \cite{Xiong2021ApproximateNN} & 0.507 & 0.295 & 0.654 & 0.296 \\
        ADORE \cite{Zhan2021OptimizingDR} & 0.514 & 0.255 & 0.590 & 0.273 \\
        RocketQA \cite{Qu2021RocketQAAO} & 0.606 & 0.319 & 0.658 & 0.316 \\
        Contriever-FT \cite{contriever} & 0.677 & 0.329 & 0.596 & 0.321 \\
        TCT-ColBERT \cite{Lin2020DistillingDR} & 0.614 & 0.316 & 0.661 & 0.309 \\
        Margin-MSE \cite{Hofsttter2020ImprovingEN} & 0.608 & 0.298 & 0.673 & 0.297 \\
        TAS-B \cite{Hofsttter2021EfficientlyTA} & 0.643 & 0.300 & 0.481 &  0.314 \\
        CLDRD \cite{Zeng:2022:CLDRD} & 0.637 & 0.348 & 0.571 & 0.327 \\ 
        \textbf{\framework (ours)} & \textbf{0.683$^*$} & \textbf{0.371$^*$} & 0.668 & 0.341$^*$ \\\midrule
        \multicolumn{5}{l}{\textbf{Some Sparse Retrieval Models (For Reference)}} \\
        BM25 \cite{Robertson1995OkapiBM25} & 0.665 & 0.236 & 0.656 & 0.299 \\
        DeepCT \cite{Dai2019ContextAwareST} & 0.630 & 0.191 & 0.406 & 0.268 \\
        docT5query \cite{Nogueira2019FromDT} & 0.675 & 0.291 & 0.713 & 0.325 \\\\
        \multicolumn{5}{l}{\textbf{Multi Vector DR Models (For Reference)}} \\
        ColBERTv2 \cite{Santhanam2021ColBERTv2EA} & \multirow{2}{*}{0.682} & \multirow{2}{*}{0.359} & \multirow{2}{*}{\textbf{0.696}} & \multirow{2}{*}{\textbf{0.357}} \\
        (DistilBERT)\\
         \bottomrule
    \end{tabular}
    \label{tab:zero-shot}
\end{table}

\paragraph{\textbf{The Zero-Shot Retrieval Results.}} All datasets used in Table~\ref{tab:results} are based on the MS MARCO passage collection and their queries are similar to that of our training set. To evaluate the model's performance under distribution shift, we conduct a zero-shot retrieval experiment on four diverse datasets: SciFact, FiQA, TREC COVID, and CQADupStack (see Section~\ref{sec:data}). In this experiment, we do not re-train any model and the ones trained on MS MARCO training set and used in Table~\ref{tab:results} are used for zero-shot evaluation on these datasets. The results are reported in Table~\ref{tab:zero-shot}. We observe that many neural retrieval models struggle with outperforming BM25 on SciFact and TREC COVID datasets. In general, the improvements observed compared to BM25 by the best performing models are not as large as the ones we observe in Table~\ref{tab:results}. This highlights the difficulty of handling domain shift by neural retrieval models. Generally speaking, the multi vector dense retrieval model (ColBERTv2) shows a more robust performance in zero-shot settings. It outperforms all single vector dense retrieval models on TREC COVID and CQADupStack. \framework performs better on the other two datasets: SciFact and FiQA. Again, we highlight that \framework has substantially lower storage requirements compared to ColBERTv2 and it also has significantly faster query processing time. Refer to Table~\ref{tab:efficiency} for more information.

\begin{table}[t]
    \centering
    \caption{Pre-retrieval query performance prediction results in terms of Pearson's $\rho$ and Kendall's $\tau$ correlations. The superscript $^\dagger$ denotes that the obtained correlations by \framework $|\Sigma_Q|$ are significant.}
    \begin{tabular}{p{2cm}cccc}\toprule
    \multirow{2}{*}{\textbf{QPP Model}} &        \multicolumn{2}{c}{\textbf{TREC-DL'19}}&        \multicolumn{2}{c}{\textbf{TREC-DL'20}}\\
    &  \textbf{P-$\rho$} & \textbf{K-$\tau$} & \textbf{P-$\rho$} & \textbf{K-$\tau$}  \\\midrule
    Max VAR \cite{Carmel:2010} & 0.138 & 0.148 & 0.230 & 0.266 \\
    Max SCQ \cite{Zhao:2008} & 0.119 & 0.109 & 0.182 & 0.237 \\
    Avg IDF \cite{Carmel:2010} & 0.172 & 0.166 & 0.246 & 0.240 \\
    SCS \cite{He:2004} & 0.160 & 0.174 & 0.231 & 0.275 \\
    Max PMI \cite{Hauff:2010} & 0.098 & 0.116 & 0.155 & 0.194 \\
    $P_{\text{clarity}}$ \cite{Roy:2019} & 0.167 & 0.174 & 0.191 & 0.217 \\
    Max DC \cite{Arabzadeh:2020} & \textbf{0.341} & \textbf{0.294} & 0.234 & 0.244 \\
    \framework $|\Sigma_Q|$ & 0.271$^\dagger$ & 0.259$^\dagger$ & \textbf{0.272$^\dagger$} & \textbf{0.298$^\dagger$}  \\
    \bottomrule
    \end{tabular}
    \label{tab:qpp}
\end{table}

\paragraph{\textbf{Exploring the Learned Variance Vectors.}} In our exploration towards understanding the representations learned by \framework, we realize that the norm of our covariance matrix for each query is correlated with the ranking performance of our retrieval model for that query. This observation motivated us to use the learned $|\Sigma_Q|$ for each query as a pre-retrieval query performance predictor (QPP). Some other well known pre-retrieval (i.e., based solely on the query and collection content, not any retrieved results) performance predictors include distribution of the query term IDF weights, the similarity between a query and the underlying collection; and the variability with which query terms occur in documents~\cite{Zhao:2008}. 

We compare our prediction against some of these commonly used unsupervised pre-retrieval QPP methods in Table~\ref{tab:qpp}. They include: 
\begin{itemize}[leftmargin=*]
    \item Max VAR \cite{Carmel:2010}: VAR uses the maximum variance of query term weight in the collection. 
    \item Max SCQ \cite{Zhao:2008}: It computes a TF-IDF formulation for each query term and returns the maximum value. 
    \item Avg IDF \cite{Carmel:2010}: This baseline uses an inverse document frequency formulation for each query term and return the average score. 
    \item SCS \cite{He:2004}: It computes the KL divergence between the unigram query language model and the collection language model.
    \item Max PMI \cite{Hauff:2010}: It uses the point-wise mutual information of query terms in the collection and returns the maximum value. 
    \item $P_{\text{clarity}}$ \cite{Roy:2019}: This baseline uses Gaussian mixture models in the embedding space as soft clustering and uses term similarity to compute the probability of each query term being generated by each cluster.
    \item Max DC \cite{Arabzadeh:2020}: This approach uses pre-trained embeddings to construct an ego network and computes Degree Centrality (DC) as the number of links incident upon the ego. 
\end{itemize}

Following the QPP literature \cite{Carmel:2010,NeuralQPP,Hauff:2010,Clarity}, we use the following two evaluation metrics: Pearson's $\rho$ correlation (a linear correlation metric) and Kendall's $\tau$ correlation (a rank-based correlation metric). We only report the results on the TREC DL datasets, since MS MARCO DEV only contains one relevant document per query and may not be suitable for performance prediction tasks. We observe that relative to existing pre-retrieval QPP approaches, \framework $|\Sigma_Q|$ has a high correlation with the actual retrieval performance. All of these correlations are significant ($p\_value < 0.05$). Note that \framework is not optimized for performance prediction and its goal is not QPP and these results just provide insights into what a model with multivariate representation may learn.

\section{Conclusions and Future Work}
This paper introduced \framework -- a novel representation learning paradigm for neural information retrieval. It uses multivariate normal distributions for representing queries and documents, where the mean and variance vectors for each input query or document are learned using large language models. We suggested a theoretically sound and empirically strong retrieval model based on multivariate Kullback-Leibler (KL) divergence between the learned representations. We showed that the proposed formulated can be approximated and used in existing approximate nearest neighbor search algorithms for efficient retrieval. Experiments on a wide range of datasets showed that \framework advances state-of-the-art in single vector dense retrieval and sometimes even outperforms the state-of-the-art multi vector dense retrieval model, while being more efficient and requiring orders of magnitude less storage. We showed that the norm of variance vectors learned for each query is correlated with the model's retrieval performance, and thus it can be used as a pre-retrieval query performance predictor. 

Multivariate representation learning opens up many exciting directions for future exploration. Given the flexibility of multivariate normal distributions, the representations learned by the model can be easily extended. For instance, linear interpolation of multivariate normals is a multivariate normal distribution. Therefore, one can easily extend this formulation to many settings, such as (pseudo-) relevance feedback, context-aware retrieval, session search, personalized search, and conversational retrieval. Furthermore, such a representation learning approach can be extended to applications beyond standard IR problems. They can be used in representation learning for users and items in collaborative filtering models, graph embedding for link prediction and knowledge graph construction, and information extraction. Another promising research direction is enhancing retrieval-enhanced machine learning (REML) models \cite{REML} using multivariate representations. MRL enables REML models to be aware of the retrieval confidence and data distribution for making final predictions. 

\section*{Acknowledgments}
This work was supported in part by the Google Visiting Scholar program and in part by the Center for Intelligent Information Retrieval. Any opinions, findings, and conclusions or recommendations expressed in this material are those of the authors and do not necessarily reflect those of the sponsors.



\end{document}